\documentstyle[12pt]{article}

\begin{document}
Cavendish preprint HEP 94/6 \\
hep-ph/9408292 \\
\vspace{1.0cm}
\begin{center}
{\huge \bf THE TRACE ANOMALY AND THE PARTON MODEL \\ }
\vspace{3ex}
\vspace{3ex}
{\large \bf S. D. Bass \\}
\vspace{3ex}
{\it Cavendish Laboratory, University of Cambridge, Madingley Road,
Cambridge CB3 0HE, U.K. \\}
\vspace{3ex}

\vspace{3ex}
{\large \bf ABSTRACT \\}
\end{center}
\vspace{3ex}
{
The trace anomaly is relevant to the renormalisation of the unpolarised
parton distributions.
The coefficient which describes the local $\gamma^{*}$(glue) interaction
differs
at next-to-leading order
between the
renormalised operator product expansion and parton model descriptions of
deep inelastic scattering
-- independent of the factorisation scheme.  }

\vspace{1.5cm}

There are two approaches to the theory of deep inelastic scattering (DIS).
These are the operator product expansion (OPE) and the QCD improved parton
model (PM) [1].
In both approaches we write the transverse and longitudinal
structure functions as the sum over the
convolution of
soft quark and gluonic parton distributions with hard coefficients, viz.
\begin{equation}
F_{[T,L]}(x,Q^2) =
\sum_q e_q^2 x (q + {\overline q}) \otimes C^q_{[T,L]} (x, \alpha_s, Q^2) +
x g \otimes C^g_{[T,L]} (x, \alpha_s, Q^2)
\end{equation}
The hard
coefficients $C^q_{[T,L]}$ and $C^g_{[T,L]}$ describe the interaction of
the hard photon (carrying transverse
or longitudinal polarisation)
with the quark and gluon partons.

In the parton model we calculate the deep inelastic cross section from a
quark or gluon target to
some order
in perturbation theory and then use the factorisation theorem to absorb
the infra-red collinear singularities
into the parton
distributions --- or, equivalently, the hadron wavefunction.
The remaining infra-red safe part of the phase space is identified with the
hard coefficients.
This factorisation is universal to all inclusive, high-energy hadronic
interactions.

Having taken care of the collinear singularities we also need to renormalise
the parton distributions.
Here we use the operator product expansion, which is applied to the total
deep inelastic cross section.
The operator product expansion
defines the target
wavefunction
in terms of the matrix elements of a tower of renormalised, local operators
[1].
These operator matrix elements need to be renormalised consistent
with the symmetry constraints of anomaly theory [2,3] --- in this case, the
trace anomaly [4].
This renormalisation has to be carried out in addition to the renormalisation
of the propagators and vertices in QCD.
In this paper I explain how the anomaly
induces a difference at next-to-leading order between the quark distributions
and gluonic coefficients
which are
obtained via the operator product expansion and the (factorisation only)
parton model.
This difference is independent of the factorisation scheme and is related to
how we describe the local photon gluon
interaction.
I first explain why the anomaly is relevant to the theory of unpolarised
deep inelastic scattering
and then show how it manifests itself in the gluonic coefficient.
Finally,
these results are compared with the axial anomaly in polarised deep inelastic
scattering.

The parton distributions in equ.(1) are defined with respect to the operator
product expansion so that
their even moments project out the target matrix elements
of the renormalised, spin-even, local operators
\begin{equation}
(p_+)^{2n+2} \int_0^1 dx \ x^{2n+1}
\biggl[ q(x, Q^2) + \overline{q}(x, Q^2) \biggr]_k =
<p | \biggl[ {\overline q}(0) \gamma_+ (iD_+)^{2n+1} {\lambda_k \over 2}
q(0) \biggr]_{Q^2}^R |p >_c
\end{equation}
and
\begin{equation}
(p_+)^{2n+2} \int_0^1 dx \ x^{2n+1} g(x, Q^2) =
<p | \biggl[ {\bf \rm Tr} G_{+ \alpha}(0) (iD_+)^{2n} G^{\alpha}_{\ +}(0)
\biggr]_{Q^2}^R |p>_c
\end{equation}
Here $G_{\mu \nu}$ is the gluon field tensor,
$D_{\mu}=\partial_{\mu}+igA_{\mu}$ is the gauge covariant derivative in QCD
and $p_{\mu}$ is the proton momentum vector.
The second ($n=0$) moment of the quark (gluon) distribution
projects out
the target matrix element of the quark (gluon) part of the
energy-momentum tensor $\theta_{\mu \nu}$ in QCD.
The operators in equs.(2,3) should be quoted with respect to some
renormalisation prescription $R$
--- here we have written them at the subtraction point $\mu^2=Q^2$.

The problem of anomalies is related to the symmetry of the Dirac sea at
infinite momentum.
When we quantise the theory about some
background mean field
(the non-perturbative vacuum)
then we construct the Fock expansion by keeping quark
and gluon quanta with momentum less than some ultra-violet cut-off $\lambda$.
High momentum (short distance) fluctuations greater than this cut-off
are absorbed into the theory as part of the definition of the mean field.
The anomaly has an analogy in classical physics where we consider a
thermodynamic system in equilibrium with a heat bath [5].
There can be a nett flow of energy between the system and the heat bath
at equilibrium
so that the {\it bare} energy of the system is not conserved.
However, we can write down a {\it free} energy, which is conserved.
We may think of the perturbative Fock space as the system and the mean
field as the heat bath.
One finds that there is a flux of energy between the mean field and the Fock
space but the
free energy remains constant [5].
The physical energy momentum tensor $\theta_{\mu \nu}$ corresponds to the
free energy.
It is conserved but picks up a finite trace due to the anomaly [6]:
\begin{equation}
\theta^{\mu}_{\mu} = (1+\gamma_m) \sum_q m_q {\overline q} q +
{\beta(\alpha_s) \over 4 \alpha_s} G_{\mu \nu} G^{\mu \nu}.
\end{equation}
Here $m_q$ is the running quark mass, $\gamma_m$ is the mass anomalous
dimension and $\beta (\alpha_s)$ is the QCD beta function.
We can regard the Fock space as a closed system (independent of the mean
field) only when we consider anomaly free quantities such as the vector
current, which measures the valence quark number.
The anomalous gluonic term in equ.(4) is particularly important
in hadronic physics because
of the contribution it makes to the mass of the proton [7,8].
The matrix element of $\theta_{\mu \nu}$ in the proton is
\begin{equation}
<p| \theta_{\mu \nu} |p> = 2 p_{\mu} p_{\nu}
\end{equation}
The trace of this equation is
\begin{equation}
<p| \theta_{\mu}^{\mu} |p> = 2 m_p^2
\end{equation}
If we were to lose the anomaly, then we would break Poinc{\'a}re invariance
at the point
of regularisation and
our parton distributions would correspond to a proton with the wrong mass.

To see how the anomaly contributes to the unpolarised structure function we
apply the operator product
expansion analysis to unpolarised photon gluon fusion at
$O(\alpha_s)$ --- Fig. 1.
Here $q_{\mu} = (q_0; 0_T, q_3)$ and $p_{\mu} = (p_0; 0_T, p_3)$ are
the
momenta of the hard photon and the soft gluon respectively;
$k$ is the momentum of the struck quark or anti-quark.
We consider the transverse cross section where the photon and gluon both
carry transverse polarisation
and follow the discussion of polarised deep inelastic scattering given in
the papers of ref.[9].
A finite
quark mass $m$ and gluon virtuality $-p^2$ is used to regulate the mass
singularity,
which arises when the struck quark carries zero transverse momentum.
The cross section for this photon gluon fusion process is:
\begin{eqnarray}
{\alpha_s \over 2\pi} \int_0^{{1\over 4} W^2}
{dk_T^2 \over \sqrt{1 - {4k_T^2 \over W^2}}}
\Biggl[& &(2x^2-2x+1) \Biggl({x \over Q^2(1-x)} +
{1 \over k_T^2 +m^2 - p^2x(1-x)} \Biggl) \\ \nonumber
&+& x(1-x) {2m^2 +p^2(2x^2-2x+1) \over (k_T^2+m^2-p^2x(1-x))^2 } \Biggr]
\end{eqnarray}
where $W^2$ is the centre of mass energy squared.
There are three distinct contributions to the total phase space.
The first
term in equ.(7)
receives contributions only from $k_T^2 \sim Q^2$, where the photon makes
a local interaction with the target gluon.
The second term receives contributions from the full range of $k_T^2$.
It is sensitive to the mass singularity at zero $k_T^2$ and gives the
$\ln Q^2$ dependence of the total cross section after we integrate over
the transverse momentum.
The third term is infra-red dominated and receives contributions only from
$k_T^2 \sim m^2, -p^2$.

At $O(\alpha_s)$ the cross section for photon gluon fusion is the sum of
the gluonic
coefficient $C^g(x, \alpha_s)$
and the quark distribution of the gluon $(q+\overline{q})^{(g)}(x, Q^2)$.
In the parton model we set the coefficients equal
to the ``hard"
part of the phase space, where
``hard" is defined to include the whole phase space which is not
subtracted
into the target wavefunction via the factorisation theorem.
The local $\gamma^{*}$(glue) interaction is clearly ``hard" in this picture
and is included into the parton model
as a part of the gluonic coefficient $C^g(x,\alpha_s)$.
The simplest way to isolate the mass singularity involves choosing a cut-off
on the transverse momentum
of the quarks
so that we identify the gluonic coefficient with the phase space
corresponding to quark jets with transverse momentum $k_T^2 > \mu^2_{fact}$,
where $\mu^2_{fact}$ is some finite scale
($Q^2 \gg \mu^2_{fact}$ and $\mu^2_{fact} \gg -p^2, m^2$).
With this cut-off, the hard parton model coefficient is:
\begin{equation}
C_{PM}^g(x,{Q^2 \over \mu^2_{fact}}, \alpha_s) =
{\alpha_s \over 2\pi} \Biggl[ (2x^2-2x+1)
\Biggl(-1 + \ln {Q^2 \over \mu^2_{fact}} + \ln {1-x \over x} \Biggr) \Biggr]
\end{equation}
The remaining soft part of the phase space is identified with the quark
distribution of the gluon
$x (q+{\overline q})_{PM}^{(g)}(x, \mu^2_{fact})$, which includes
all quarks and
anti-quarks with transverse momentum less than $\mu^2_{fact}$:
\begin{eqnarray}
& &(q + \overline{q})^{(g)}_{PM}(x, \mu^2_{fact}) \\ \nonumber
&=&{\alpha_s \over 2 \pi}
\biggl[ (2x^2-2x+1) \ln {\mu^2_{fact} \over m^2 - p^2 x(1-x)}
+ x(1-x) {2m^2 + (2x^2-2x+1)p^2 \over m^2 - p^2 x(1-x)} \biggr]
\end{eqnarray}
We now compare this parton model distribution with the renormalised
quark
distribution,
which we calculate via the operator product expansion.

Working in the light cone gauge ($A_+=0$),
the quark distribution is the renormalised version of the graph in Fig. 2:
\begin{eqnarray}
& &4p_+^2 x (q+ \overline{q})^{(g)}(x) = \\ \nonumber
& & g^2 \int {d^4k \over (2 \pi)^4} \delta(x - {k_+ \over p_+})
(2 \pi) \delta ((k-p)^2 - m^2)
{ {\rm Tr} \biggl[ (\hat{k} +m) \hat{\epsilon}^* (\hat{k}-\hat{p}+m)
\hat{\epsilon} (\hat{k} + m) \gamma_+ k_+ \biggr]
\over (k^2 - m^2 -i \epsilon)^2 }
\end{eqnarray}
The $k_+$ and $k_-$ integrals are carried out using the two delta function
constraints
to yield an ultra-violet divergent integral in the Euclidean $k_T^2$, which
needs to be renormalised.
We can choose to regularise this ultra-violet divergence via dimensional
regularisation
or the introduction of a Pauli-Villars term -- both of which respect the
anomaly.
If we use dimensional regularisation, then we continue the physics in the
transverse dimensions
from $D = 2$ to
$D= (2 - 2 \epsilon)$ dimensions;
the ultra-violet divergence appears as a term proportional to
${1 \over \epsilon}$,
which is subtracted out as the infinite renormalisation constant of the parton
distribution.
The sum over the transverse polarisations of the gluon is
\begin{equation}
\sum_{\lambda=1,2} \epsilon_{\mu}(\lambda) \epsilon_{\nu}^{*}(\lambda)
= {g_{\mu \nu}^T \over D}
\end{equation}
Equ.(10) evaluates as
\begin{equation}
4x (p_+)^2 2 \alpha_s \int_0^{\infty}
{d^{2-2\epsilon}k_T \over (2 \pi^2)^{2-2\epsilon} }
\biggl[ {(2x^2 -2x +1) + {2 \epsilon \over D} 2 x(1-x)
\over k_T^2 + m^2 - p^2 x(1-x)}
+ x(1-x) {2m^2 + (2x^2-2x+1)p^2 \over [k_T^2 + m^2 -p^2 x(1-x)]^2 }
\biggr]
\end{equation}
in $D = (2-2\epsilon)$ dimensions.
If no ultraviolet
cutoff is imposed, then the integral
\begin{equation}
\int d^{2-2\epsilon}k_T { 1 \over k_T^2 + m^2 - p^2 x(1-x)}
\end{equation}
develops an ultra-violet pole ${1 \over \epsilon}$,
which comes from $k_T^2$ much greater than the parton model cut-off
$\mu^2_{fact}$.
This pole
cancels with the $\epsilon$ term in the numerator of equ.(12)
to yield
a finite term which is induced by the regularisation:
the (trace) anomaly
in the energy-momentum tensor.
This anomalous term is independent of how we next make the minimal
subtraction
to obtain the renormalised parton distribution.

In the $\overline{MS}$ scheme the renormalised quark distribution is
\begin{equation}
(q+\overline{q})^{(g)}_{R}(x, \mu^2) =
(q+\overline{q})^{(g)}_{PM}(x, \mu^2) + C^g_A(x, \alpha_s)
\end{equation}
where
$C^g_A$ is the anomaly at $O(\alpha_s)$
\begin{equation}
C^g_A (x, \alpha_s) = {\alpha_s \over \pi} x (x-1)
\end{equation}
and $\mu^2$ is the renormalisation scale.
Here we have subtracted out the divergent term
${\alpha_s \over 2\pi} (2x^2-2x+1) ({1 \over \epsilon} - \gamma_E + \ln 4\pi)$
into the renormalisation constant.
In the general minimal subtraction scheme we subtract out a divergent
term
proportional to $({1 \over \epsilon} + a)$, where $a$ is some constant;
the renormalised quark distribution is equal to the parton model distribution
in equ.(9) with $\mu^2_{fact} = \mu^2 e^{\ln 4\pi - \gamma_E - a}$
plus the subtraction independent
anomaly, which is
left behind
by the ultra-violet regularisation.
Pauli-Villars regularisation yields the same
renormalised quark distribution as equs.(14,15).
In this case, the anomaly comes from $k_T^2$ of order the mass squared of
the Pauli-Villars fermion
in the second term of equ.(12).

The renormalised quark distribution corresponds to a gluonic coefficient
which is different to the parton model by the anomalous term
in equ.(15);
viz.
\begin{equation}
C^g_{OPE}(x,{Q^2 \over \mu^2_{fact}}, \alpha_s) =
C_{PM}^g(x,{Q^2 \over \mu^2_{fact}}, \alpha_s)  - C^g_A (x, \alpha_s)
\end{equation}
Since the anomaly term $C^g_A (x, \alpha_s)$ is a polynomial in $x$ at
$O(\alpha_s)$,
it describes a local $\gamma^{*}$(glue) interaction.
Some of this
local interaction
is included as a contact interaction with the target wavefunction within the
OPE approach,
which offers the most rigorous treatment of the total deep inelastic
cross-section.
The parton model
treatment of the local $\gamma^{*}$(glue) interaction is complete
if one can treat
the proton target as an ensemble of perturbative quark and gluon degrees
of freedom.
However, it is not clear ab initio that the parton model ``scheme" is
sufficiently general
that it can describe all local interactions between the hard photon
and the background
colour mean field
in addition to the perturbative glue.
In particular, it is not clear that the former interaction has the same
two-quark-jet
signature that we expect from photon gluon fusion in perturbation theory.
The background field via the anomaly is important in determining the mass
of the proton, equ.(6).
We need to understand how it matches onto the perturbative glue.
The factorisation theorem tells us how to treat the infra-red collinear
singularities
but does not tell us how to treat the local
$\gamma^{*}$(glue) interaction, which is strictly non-collinear.

The parton model gluonic coefficient is recovered from the operator product
expansion analysis if we define our parton model quark
distribution
to be
\begin{equation}
x(q + {\overline q})_{PM}(x, Q^2) \equiv
x(q + {\overline q})_{R}(x, Q^2) - x g_R \otimes C^g_A (x,Q^2)
\end{equation}
Note
that this $x(q + {\overline q})_{PM}(x, Q^2)$
is not the same object
that we obtain with a fixed ultra-violet cut-off on the
transverse momentum.
The $k_T^2$ cut-off breaks Poincar{\'e} invariance at the point of
regularisation [2,3,5] (leading to a non-conserved $\theta_{\mu \nu}$),
whereas the quark distribution on the left hand side of equ.(17) is the
difference of two renormalised quark distributions ---
each of which includes the effect of the anomaly at infinite momentum.

In the parton literature it is commonly assumed that the parton model
distributions
can be written as light-cone correlation functions
[10,11]
\begin{eqnarray}
q(x) &=& {1 \over 2\pi} \int dz_{-} e^{-ixz_- p_+}
<p| \overline{q}(z_-) \gamma_+ {\cal P} e^{ig \int_0^{z_-} dy_- A_+} q(0)|p>_c
\\ \nonumber
{\overline q}(x) &=& -{1 \over 2\pi} \int dz_{-} e^{ixz_- p_+}
<p| \overline{q}(z_-) \gamma_+ {\cal P} e^{ig \int_0^{z_-} dy_- A_+} q(0)|p>_c.
\end{eqnarray}
(The moments of these correlation functions
project out the target matrix elements of the operators in the OPE.)
It is clear from our discussion
that this identification needs to be modified according to equ.(17)
because
of the anomaly.

The correlation function offers
a simple, physical picture of the anomaly in deep inelastic
scattering.
In a quark model (which has an inherent ultra-violet cut-off)
the
non-local matrix element in equ.(18) has the interpretation that we
take out
a quark (insert an anti-quark) in the nucleon and re-insert it (take it out)
at position $z_-$ along
the light-cone.
The $z_- \rightarrow 0$ limit of this matrix element
is
the same object that appears in Schwinger's derivation of the anomaly,
which uses point splitting [12] (along the light-cone).
Intuitively, one can think of the anomaly in deep inelastic scattering
as a zero correlation length effect
which is missing
in the parton model.

We now compare our results with the anomaly in polarised deep inelastic
scattering.
The anomaly contribution to the renormalised, unpolarised quark
distribution (equ.(14))
arises
in the same way as the axial anomaly contribution to the spin dependent
quark distribution [9],
which is measured in polarised deep inelastic scattering.
The anomaly comes from $k_T^2$ much greater than the parton model cut-off.
Clearly, the theory of polarised and unpolarised deep inelastic scattering
should be formulated in a consistent way.
If we imposed some ultra-violet cut-off on the $k_T^2$ then we would lose the
trace anomaly
in the unpolarised structure function and also the axial anomaly in the
polarised
structure function.
This means that the EMC spin effect [13]
and the physics of the unpolarised structure function have to be discussed
together as part of the same problem.

The renormalised spin dependent quark and gluon distributions which are
measured in polarised deep inelastic scattering
are
\begin{equation}
2M s_+ (p_+)^{2n} \int^1_0 dx \ x^{2n} \Delta q_{k} (x, Q^2) =
<p,s | \biggl[ {\overline q}(0) \gamma_+ \gamma_5 (i D_+)^{2n}
{\lambda^k \over 2}
q(0) \biggr]_{Q^2}^{R} |p,s >_c
\end{equation}
\begin{equation}
2M s_+ (p_+)^{2n} \int^1_0 dx \ x^{2n} \Delta g (x, Q^2) =
<p,s | \biggl[ {\bf \rm Tr} \ G_{+ \alpha}(0) (iD_+)^{2n-1}
{\tilde G}^{\alpha}_{\ +}(0) \biggr]_{Q^2}^{R} |p,s >_c
(n \geq 1)
\end{equation}
where
${\tilde G}_{\mu \nu} = {1 \over 2} \epsilon_{\mu \nu \alpha \beta}
G^{\alpha \beta} $ is the dual tensor and
$s_{\mu}$
denotes the proton spin vector ($M$ is the proton mass).
The spin dependent distributions have to be renormalised consistent the
axial anomaly,
which says that any ultra-violet regularisation that respects gauge invariance
does not respect chirality as a good symmetry.
The renormalisation of each of the C-even axial tensors in equ.(19) involves
a gauge dependent counterterm $k_{\mu \mu_1 ... \mu_{2n}}$.
These $k_{\mu \mu_1 ... \mu_{2n}}$ define a gauge dependent gluonic
distribution
$\kappa (x,Q^2)$ [14].

There is an important physics difference between the anomalies in unpolarised
and polarised deep inelastic scattering.
In the unpolarised case the physical energy-momentum tensor is conserved: the
canonical identification
of $x(q+{\overline q})_R$ with a quark momentum distribution holds true in the
renormalised theory.
The physical axial vector current is not conserved [15] and has a two-loop
anomalous dimension, which
is induced by the anomaly [16] --- it does not measure spin [14,17].
This means that $\Delta q_R(x,Q^2)$
does not have a canonical interpretation.
The colour mean field in the proton should be manifest explicitly
in $\Delta q_R(x,Q^2)$.
This effect may be seen as an OZI violation in the spin dependent structure
function $g_1$ at large $x$ [14].

Just as in the unpolarised case (equ.16)), the renormalised quark distribution
in equ.(19)
yields a spin dependent gluonic coefficient which differs from the parton
model.
The parton model coefficient [18] is recovered if we define
\begin{equation}
\Delta q_{PM}(x,Q^2) = \Delta q_{R}(x,Q^2) - \Delta g_R \otimes {\tilde C}^g_A
(x, Q^2)
\end{equation}
where ${\tilde C}^g_A$ is the anomalous coefficient at $O(\alpha_s)$.
It is important to distinguish the $\Delta g_R(x,Q^2)$ in equ.(21)
from the gauge dependent $\kappa_R (x,Q^2)$.
In the light-cone gauge $A_+=0$ (where the parton model is usually formulated)
one finds
$\kappa_R(x,Q^2) = \Delta g_R \otimes {\tilde C}^g_A (x,Q^2)$.
However, in a
covariant gauge
$\kappa_R (x,Q^2)$ is not invariant under gauge transformations ---
in contrast to
the physical $\Delta q_R (x,Q^2)$ and $\Delta g_R (x,Q^2)$ [14].
In polarised deep inelastic scattering the axial anomaly introduces a
second
set of (gauge dependent) gluonic operators.
In unpolarised deep inelastic scattering the trace anomaly involves
the same leading twist operators that define the regular gluon distribution
in equ.(3) in order to yield
the trace
in equ.(4)
--- see also [5].

In summary,
the anomaly means that the local $\gamma^{*}$(glue) interaction is treated
differently in the operator product expansion and parton model treatments
of both unpolarised
and polarised deep inelastic scattering.
Since an operator analysis of anything other than the total deep inelastic
cross-section is
intractable,
there is no clear way of renormalising the initial and final state wavefunction
in other high-energy collisions.
It is interesting to consider the possibility that the anomaly might induce a
process dependence in
the parton distributions
at greater than or equal to next-to-leading order in $\alpha_s$.

\vspace{2.0cm}

{\bf Acknowledgements}

I thank N. N. Nikolaev, F. Steffens, A. W. Thomas and B. R. Webber
for helpful
discussions.
It is a pleasure to acknowledge the hospitality of the theoretical
physics group at the University of Adelaide,
where part of this work was completed.
This work was supported in part by the UK Particle Physics and Astronomy
Research Council
and the EU Programme ``Human Capital and Mobility", Network
``Physics at High Energy Colliders", contract CHRX-CT93-0537 (DG 12 COMA).

\pagebreak

{\bf References}
\vskip 12pt
\begin{enumerate}
\item
R. G. Roberts, {\it The structure of the proton}, Cambridge UP (1990)
\item
P. A. M. Guichon, l'Aquila lectures, Saclay preprint DPhN-Saclay-9136 (1990)
\item
J. Zinn-Justin, {\it Quantum field theory and critical phenonema},
Chapter 7, Oxford UP (1989)
\item
R. J. Crewther, Phys. Rev. Letts. 28 (1972) 1421 \\
M. S. Chanowitz and J. Ellis, Phys. Letts. B40 (1972) 397;
Phys. Rev. D7 (1973) 2490
\item
V. N. Gribov, Budapest preprint KFKI-1981-66 (1981)
\item
M. Shifman, Phys. Rep. 209 (1991) 341
\item
M. A. Shifman, A. I. Vainshtein and V. I. Zakharov,
Nucl. Phys. B147 (1979) 385, 448
\item
R. L. Jaffe and A. Manohar, Nucl. Phys. B337 (1990) 509
\item
R. D. Carlitz, J. C. Collins and A. H. Mueller, Phys. Lett. B214 (1988) 229 \\
G. T. Bodwin and J. Qiu, Phys. Rev. D41 (1990) 2755 \\
S. D. Bass, B. L. Ioffe, N. N. Nikolaev and A. W. Thomas, J. Moscow Phys. Soc.
1 (1991) 317 \\
G. G. Ross and R. G. Roberts, Rutherford preprint RAL-90-062 (1990)
\item
J. C. Collins and D. E. Soper, Nucl. Phys. B194 (1982) 445
\item
C. H. Llewellyn Smith, Oxford preprint OX-89/88 (1988)
\item
J. Schwinger, Phys. Rev. 82 (1951) 664
\item
for reviews see: \\
G. Veneziano, Okubofest lecture, CERN TH 5840 (1990) \\
S. D. Bass and A. W. Thomas, Prog. Part. Nucl. Phys 33 (1994) 449
\item
S. D. Bass, Z Physik C55 (1992) 653, C60 (1993) 343
\item
J. S. Bell and R. Jackiw, Nuovo Cimento A51(1969) 47 \\
S. L. Adler, Phys. Rev. 177 (1969) 2426
\item
J. Kodaira, Nucl. Phys. B165 (1980) 129
\item
G. Veneziano, Mod. Phys. Letts. A4 (1989) 1605
\item
A. V. Efremov and O. V. Teryaev, Dubna preprint E2-88-287 (1988) \\
G. Altarelli and G. G. Ross, Phys. Letts. B212 (1988) 391
\end{enumerate}
\end{document}